# Magnetic moment of the pentaquark $\Theta^+$ state

A. R. Haghpayma[†]

*Department of Physics, Ferdowsi University of Mashhad*

*Mashhad, Iran*

## Abstract

Although the $\Theta^+$ has been listed as a three star resonance in the 2004 PDG, its existence is still not completely establisheed, Whether the $\Theta^+$ exist or not, but it is still of interest to see what QCD has to say on the subject. The baryon magnetic moment is a fundamental observable as its mass which encodes information of the underlying quark gluon structure and dynamics. Assuming a conventional correlated perturbative chiral quark model (CP$\chi$QM), we suggest that the $\Theta^+$ baryon is a bound state of two vector diquarks and a single antiquark, the spatially wave function of these diquarks has a P - wave and a S - wave in angular momentum in the first and second version of our model respectively, as the result of these considerations we construct the orbital colour - flavour - spin symmetry of $q^4\bar{q}$ contribution of quarks. Then we calculate the $\Theta^+$ magnetic moment in our model.

## 1. Introduction

The year 2003 may be remembered as a renaissance of hadron spectroscopy, at the early's of that year ( LEPS ) collaboration T. Nakano et al.[1] reported the first evidence of a sharp resonance $Z^+$ renamed to $\Theta^+$ at $M_{\Theta} \simeq 1,54 \pm 0,01$ Gev with a width smaller than $\Gamma_{\Theta^+} < 25$ Mev.

The experiment performed at the Spring - 8 facility in Harima japan and this particle was identified in the $K^+N$ invariant mass spectrum in the photo - production reaction $\gamma n \to K^- + \Theta^+$, which was induced by a Spring - 8 tagged photon beam of energy up to 2.4 Gev.

The existence of $\Theta^+$ was soon confirmed by various groups in several photo — nuclear reactions[2] including V. V. Barmin etal.[3] ITEP( DIANA )[4] JLAB[1]( CLAS )[5] and ELSA( SAPHIR )[10].

This discovery has triggered an intense experimental and theoretical activity to understand the structure of the state.

Such states are believed to belong to a multiplet of states where the possible observability of the other members has to be worked out.

With the conventional constituent quark model, the conservation rules guarantee that it has a strangeness S=1, baryon number B=1 and charge Q = 1, thus the hypercharge is Y = B + S = 2 and the third component of isospin is I= 0 . no corresponding $pK^+$ ( I=1 ) state is observed at the same mass, due to absence of a $\Theta^{++}$ in the $\gamma p \to pK^+K^-$ channel and thus the isospin of $\Theta^+$ is the same I =0 and it also seems important that no S =1 baryonstates have been observed below the NK threshold and this state seems to be the ground state.

All known baryons with B = 1 carry negative or zero strangeness a baryon with strangeness S = 1, it should contain at least one $\bar{s}$, can not consist of three quarks, but must contain at least four quarks and an antiquark ; in other words, must be a pentaquark or still more complicated object. now its called $\Theta^+$ pentaquark in literature.

From the charge and the strangeness. $u^2 d^2 \bar{s}$ is a possibility as the content of $\Theta^+$, which called the minimum quark content, such state is exotic; in general states with the $\bar{q}$ having different flavour than the other four quarks and whose quantum numbers cannot be defined by 3 quarks alone are called exotics. thus we have an exotic $\Theta^+$.

The mass and the width of $\Theta^+$ and other exotic pentaquark baryons has predicted by several hadron models. its width ( $\Gamma < 10$ Mev ) is exceptionally narrow as for a hadron resonance located at 110 Mev above the NK threshold usually refered to narrow width puzzle.[6]

There is no direct measurement of its spin S and Isospin I and its angular momentum J and parity P are different in various theoretical works, however most of them postulated its angular momentum J to be J = 1/2 but the possibility of J = 3/2 and S =1/2 and P = + is rather plausible.

## 2. Pentaquark As A Bound State of Two Vector Diquarks and One Anti-quark

Assuming a conventional correlated perturbative chiral quark model (CP$\chi$QM), we suggest that the $\Theta^+$ baryon is a bound state of two diquarks and a single antiquark, the spatially wave function of these diquarks has a p - wave and a s - wave in angular momentum in the first and second version of our model [7] respectively.

The $[2]^f$ flavour symmetry of each diquark leads to $[22]^f_{\bar{t}}$ flavour symmetry for $q^4$, the $[2]^s$ spin symmetry of each diquark leads to $[22]^s$ and $[31]^s$ spin symmetry for $q^4$ in the first version and second version of our model respectively.

The colour symmetry of each diquark is $[11]^c$ and for the first version of our model we assume $[2]^c$ for one of the diquark pairs this leads to $[211]^c$ color symmetry for $q^4$.

The orbital symmetry of each diquark is $[2]^o$ and for the first version we assume $[11]^o$ for one of the diquark pairs, this leads to $[31]^o$ and $[4]^o$ orbital symmetry for $q^4$ in the first and second version of our model respectively.

Briefly the spin - flavour - colour and parity of our model for the first version and second one are as follows:

$$\left| (QQ)^{\ell=1,3_c,\bar{6}_f} \bar{q}^{j=\frac{1}{2},\bar{3}_c,\bar{3}_f} \right\rangle^{J^\Pi = (\frac{1}{2}^+ \oplus \frac{3}{2}^+), \mathbf{1}_c, (\overline{\mathbf{10}}_f \oplus \mathbf{8}_f)} \quad (1)$$

$$\left| (QQ)^{\ell=0,3_c,\bar{6}_f} \bar{q}^{j=\frac{1}{2},\bar{3}_c,\bar{3}_f} \right\rangle^{J^\Pi = (\frac{1}{2}^- \oplus \frac{3}{2}^-), \mathbf{1}_c, (\mathbf{8}_f \oplus \overline{\mathbf{10}}_f)} \quad (2)$$

We have considered $[4]^{fs}_{126}$ and $[31]^{fs}_{210}$ for the flavour - spin configurations of $q^4$ in the first and second version of our model respectively, this leads to $[51111]_{700}$ and $[42111]_{1134}$ for the flavour - spin configurations of $q^4\bar{q}$, but if one assume the angular momentum $\ell = 1$ for the four quarks $q^4$ there are several allowed $SU_{fs}(6)$ representations for $q^4\bar{q}$ which are $[51111]_{700} [42111]_{1134}$ $[33111]_{560}, [32211]_{540}$ based on $[4]^{fs}, [31]^{fs}, [22]^{fs}$ and $[211]^{fs}$ $SU_{fs}(6)$ representations for $q^4$ respectively.

---

[†] e-mail: haghpeima@wali.um.ac.ir

[1] Although the JLab new experiment[8] searched for pentaquarks in the same channel at a level at least one order of magnitude better-than the previous published result and found no pentaquarks., in general, the negative reports which involve higher energies and statistics or reactions different from those that produced positive results may not directly contradict them.

The magnetic moment is an internsic observable of particles which may encode important imformation of its quark gluon structure and underlying dynamics.

Different magnetic moments will affect both the total and different cross sections in the photo - or electro production of pentaquarks. Hence, knowledge of the pentaquark magnetic moments will help us unviel the mysterious curtain over the pentaquarks at present and deepen our understanding of the underlying quark structure and dynamics.

The pentaquark magnetic moments in several typical models have been calculated.[9] now we calculate it for our model.

For the magnetic moment of a particle we have:

$$\vec{\mu} = g\vec{S} \quad (3)$$

Where $\vec{\mu}$ is magnetic moment, $g$ is gyromagnetic ratio and $\vec{S}$ is spin operator, this leads to $\mu_z = g\, S_z$, for the quarks we have:

$$g_q = g_s \mu_q = 2\mu_q = 2\frac{Q_q}{2m_q} = \frac{Q_q}{m_q} \quad (4)$$

in which $\mu_q$ is quark magneton, and $Q_q$, $m_q$ are quarks charge and mass respectively.

If the particle has angular momentum $\vec{l}$ the magnetic moment would be:

$$\vec{\mu} = g\vec{S} + g_l \vec{l} \quad (5)$$

We conclude that for the pentaquark we have:

$$\mu_z = <\psi_{fs} | \sum_i g_i S_z^i + g_{l_i} l_z^i | \psi_{fs}> \quad (6)$$

in which $\psi_{fs}$ is the flavour-spin wave function of the pentaquark.

For the second term of (6) we have:

$$\mu_z = <\psi_{fs} | \sum_i g_{l_i} l_z^i | \psi_{fs}> = <\psi_{fs} | \sum_i g_{l_i} l_z^i | \psi_{fs}>^{diquark1}$$
$$+ <\psi_{fs} | \sum_i g_{l_i} l_z^i | \psi_{fs}>^{diquark2} + <\psi_{fs} | g_{l_{\bar{s}}} l_z^{\bar{s}} | \psi_{fs}>^{\bar{s}} \quad (7)$$

The contribution of the $\bar{s}$ - term would be zero and for the first and second term we have:

$$1.\text{term} + 2.\text{term} = <\mu_1^1 l_z^1>^{diquark1} + <\mu_2^2 l_z^2>^{diquark2} = <\mu_l l_z>^{relative} \quad (8)$$

$$\mu_l = \frac{m_2 \mu_1}{m_1 + m_2} + \frac{m_1 \mu_2}{m_1 + m_2}.$$

Where $m_1, \mu_1$ and $m_2, \mu_2$ are the masses and magnetic moments for the first and second diquarks respectively.

For the first version of our model in which the two diquarks are in $l = 1$ P-wave we have:

$$\mu_l <l_z>^{relative} = \mu_l \langle 11\tfrac{1}{2} - \tfrac{1}{2} | \tfrac{1}{2}\tfrac{1}{2}\rangle^2 \quad (9)$$

But for the second version of our model in which the two diquarks are in $l = 0$ S-wave we have:

$$\mu_l <l_z>^{relative} = 0 \quad (10)$$

The contribution of $\bar{s}$ in the first term of Eq (6) is:

$$\mu_{\bar{s}} \langle 10\tfrac{1}{2}\tfrac{1}{2} | \tfrac{1}{2}\tfrac{1}{2}\rangle^2 \quad (11)$$

and for the first version of our model we have:

$$\mu_{\bar{s}} \langle 10\tfrac{1}{2}\tfrac{1}{2} | \tfrac{1}{2}\tfrac{1}{2}\rangle^2 - \mu_{\bar{s}} \langle 11\tfrac{1}{2} - \tfrac{1}{2} | \tfrac{1}{2}\tfrac{1}{2}\rangle^2 \quad (12)$$

for the contribution of $\bar{s}$ in the first term of Eq (6) due to relative angular momentum between s-quark and diquarks.

## References

[1] T. Nakano et al., Phys. Rev. Lett. **91**, 012002 (2003).

[2] S.Stepanyan et al., CLAS Collaboration, Phys. Rev. Lett. **91**, 252001 (2003), hep-ex/0307018; V. V. Barmin et al., DIANA Collaboration, Phys. Atom. Nucl. **66**, 1715 (2003) [Yad. Phys. **66**, 1763 (2003)], hep-ex/0304040; J. Barth et al., SAPHIR Collaboration, Phys. Lett. **B572**, 127 (2003), hep-ex/0307083; A. E. Asratyan et al., hep-ex/0309042, to be published in Phys. Atom. Nucl.; V. Kubarovsky et al., CLAS Collaboration, Phys. Rev. Lett. **92** (2004) 032001; A. Airapetian et al., HERMES Collaboration, hep-ex/0312044; A. Aleev et al., SDV Collaboration, hep-ex/0401024.

[3] V.V. Barmin et al., Phys. Atom. Nucl. **66**, 1715 (2003); Yad. Fiz. **66**, 1763 (2003).

[4] DIANA collaboration, V.V. Barmin, et al., Phys. Atom. Nucl **66** (2003), 1715.

[5] CLAS collaboration, S. Stepanyan, et al., Phys. Rev. Lett **91** (2003), 252001.

[6] A. R. Haghpayma, hep-ph/0606214 V1 Jun 2006.

[7] A. R. Haghpayma, hep-ph/0606162 V1 Jun 2006.

[8] CLAS collaboration, B . Mckinnon ,et al ,arXiv:hep-ex/0603028 V1 14 Mar 2006.

[9] Y.-R.Liu et al, hep-ph/0312074.

[10] J. Barth et al., [SAPHIR] Collaboration, arXiv:hep-ex/0307083.